\documentclass[reprint,superscriptaddress,amsmath,amsfonts,amssymb,aps,prl,showkeys]{revtex4-2}

\usepackage{graphicx}
\usepackage{mathtools}
\usepackage{bm}
\usepackage[makeroom]{cancel}
\usepackage{amssymb}
\usepackage{dcolumn}
\usepackage{subcaption}
\usepackage{fullpage}
\usepackage{amsmath}
\usepackage{multirow}
\usepackage{physics}
\usepackage{dcolumn}
\usepackage{xcolor}
\usepackage{upgreek}
\usepackage[normalem]{ulem}

\usepackage[figurename=Fig.]{caption}
\usepackage{graphicx}
\usepackage{caption}
\begin{document}
\title{Manipulation of Spin Transport in Graphene/Transition Metal Dichalcogenide Heterobilayers upon Twisting}

\author{Armando Pezo}
\affiliation{Center of Natural and Human Sciences, Federal University of ABC, Santo André, SP 09210-580, Brazil}
\affiliation{Catalan Institute of Nanoscience and Nanotechnology (ICN2), CSIC and The Barcelona Institute of Science and Technology, Campus UAB, 08193 Bellaterra, Catalonia, Spain}

\author{Zeila Zanolli}
\affiliation{Dept. of Chemistry, Debye Institute for Nanomaterials Science, and ETSF, Utrecht University, The Netherlands}
\affiliation{Catalan Institute of Nanoscience and Nanotechnology (ICN2), CSIC and The Barcelona Institute of Science and Technology, Campus UAB, 08193 Bellaterra, Catalonia, Spain}

\author{Nils Wittemeier}
\affiliation{Catalan Institute of Nanoscience and Nanotechnology (ICN2), CSIC and The Barcelona Institute of Science and Technology, Campus UAB, 08193 Bellaterra, Catalonia, Spain}

\author{Pablo Ordej\'on}
\affiliation{Catalan Institute of Nanoscience and Nanotechnology (ICN2), CSIC and The Barcelona Institute of Science and Technology, Campus UAB, 08193 Bellaterra, Catalonia, Spain}

\author{Adalberto Fazzio}
\affiliation{Center of Natural and Human Sciences, Federal University of ABC, Santo André, SP 09210-580, Brazil}
\affiliation{Brazilian Nanotechnology National Laboratory/CNPEM, Campinas, SP13083-970, Brazil}

\author{Stephan Roche}
\affiliation{Catalan Institute of Nanoscience and Nanotechnology (ICN2), CSIC and The Barcelona Institute of Science and Technology, Campus UAB, 08193 Bellaterra, Catalonia, Spain}
\affiliation{Institucio Catalana de Recerca i Estudis Avan¸cats (ICREA), 08010 Barcelona, Spain}

\author{Jose H. Garcia}
\affiliation{Catalan Institute of Nanoscience and Nanotechnology (ICN2), CSIC and The Barcelona Institute of Science and Technology, Campus UAB, 08193 Bellaterra, Catalonia, Spain}

\date{\today}

\begin{abstract}
Proximity effects between layered materials trigger a plethora of novel and exotic quantum transport phenomena. Besides, the capability to modulate the nature and strength of proximity effects by changing crystalline and interfacial symmetries offers a vast playground to optimize physical properties of relevance for innovative applications. In this work, we use large-scale first principles calculations to demonstrate that strain and twist-angle strongly vary the spin-orbit coupling in graphene/transition metal dichalcogenide heterobilayers. Such a change results in a modulation of the spin relaxation times by up to two orders of magnitude.  Additionally, the relative strengths of valley-Zeeman and Rashba spin-orbit coupling can be tailored upon twisting, which can turn the system into an ideal Dirac-Rashba regime or generate transitions between topological states of matter. These results shed new light on the debated variability of spin-orbit coupling and clarify how lattice deformations can be used as a knob to control spin transport. Our outcomes also suggest complex spin transport in polycrystalline materials, due to the random variation of grain orientation, which could reflect in large spatial fluctuations of  spin–orbit coupling fields.

\end{abstract}

\maketitle

\section{Introduction}

The intrinsically small spin-orbit coupling (SOC) and high mobility of graphene, allows spin relaxation lengths to reach macroscopic scales and promotes it as an ideal candidate for passive spintronics \cite{ferrari2015nanoscale,Sierra2021, Roche_2015, Hallal_2019}. Although such a small SOC  is detrimental to electrical spin control and excludes the possibility of graphene-only spintronic devices, the two-dimensional (2D) nature of graphene facilitates enhancing the SOC artificially via proximity to high SOC substrates or other 2D materials \cite{Avsar2014NC,Gmitra2016,AlsharariPRB2016,Zanolli2018,Song2018,TarikPPRB2018,Alsharari2018,Li2019,David2019,Hallal_2019,saveroMRS2020}. Graphene/transition metal dichalcogenide (TMD) heterostructures constitute a paradigmatic example of such control. In these systems, different groups measured a variety of spin-orbit-driven phenomena such as the giant spin-lifetime anisotropy \cite{AronPRL2017, Benitez2018}, spin Hall and Rashba-Edelstein effects \cite{Safeer2019,Ghiasi2019,Benitez2020,HerlingADM2020,franz_casanova}, and weak anti-localization \cite{Wang2015,BYangPRB2017,Zihlmann2018,WakamuraTPRL2018,WakamuraTPRB2019}, providing strong evidences of proximity-induced SOC and bringing SOC engineering to a reality. Nevertheless, there are still discrepancies between theory and measurements regarding the strength and mechanisms underlying proximity effects. First-principles simulations established a range from 0.1 to 1 meV for the spin-orbit coupling parameters \cite{Avsar2014NC,Gmitra2016,AlsharariPRB2016,Song2018,TarikPPRB2018,Alsharari2018,Li2019,David2019}, while the upper limit of experimental estimates can be as high as 10 meV \cite{Wang2015,WakamuraTPRL2018, WakamuraTPRB2019}. Interestingly, these effects were only reported in exfoliated systems and remain to date elusive for polycrystalline samples, maybe due to the presence of grain boundaries  \cite{Cummings2019}. An alternative explanation relies on the moiré physics of graphene on TMDs, which could lead to significant SOC modulations \cite{Alsharari2018,Li2019,David2019} but that under certain conditions could be deleterious to the proximity-induced SOC. 

A twist-angle-modulated spin-orbit coupling enriches the field of moiré physics and opens the door for novel interplays \cite{Cao2018,PhysRevLett.107.216602,Carr2020,Yankowitz1059,Wang2020,PhysRevLett.121.266401,Giustino_2021}. First-principle methods, when applied in crystals, exploit the spatial periodicity to make the calculation feasible. But computing proximity effects in incommensurate systems with ab-initio methods becomes extremely challenging, as it can lead to using prohibitive large unit cells or unrealistic strain values. Therefore, most studies limit themselves to investigate small, and aligned commensurate structures \cite{Gmitra2016} and complemented their {\it{ab initio}} results with tight-binding methods, computing at a reduced cost the effect of rotations and strain \cite{Alsharari2018,Li2019, David2019}. However, these approaches ignore the impact of the underlying symmetries and quantum mechanics effects such as charge transfer and hybridization due to the structures' relaxation \cite{PhysRevB.65.075411}. 

Here, we use large-scale DFT to unveil the SOC dependence of strain and twist-angle in graphene/TMD heterobilayers. The microscopic SOC parameters are obtained from a symmetry-based Dirac Hamiltonian fitted against the DFT band structure and spin textures. Following this procedure, we show that Rashba and valley-Zeeman interactions are highly sensitive to  strain and twist angle, which permit to tune the system into different regimes characterized by a different spin-lifetime anisotropy.  We also demonstrate that we can tune the system into an ideal Dirac-Rashba state via strain deformations. Such a state, characterized solely by the Rashba SOC parameters, has not yet been reported in similar systems and could lead to strong spin-orbit torque components of particular interest for low-power memory applications \cite{RMPManchon_2019,Manchon2015_sot}. Finally, we compute the topological invariant under different conditions and demonstrate that a twist angle can induce topological transitions by modulating the Kane-Mele spin-orbit coupling.

\section{Models and Methods}

To evaluate the strain and twist-angle effects, we considered different graphene/TMD heterostructures constructed using $N\times N$ TMD and $M\times M$ graphene primitive cells, which we denoted as $N\times M$ through the manuscript. Then, on each of these cells we determined the band structure and spin textures by employing fully relativistic density functional theory (DFT) as implemented in the SIESTA code \cite{Soler2002,siesta2}. We describe the SOC within a fully relativistic pseudo-potential formulation \cite{Cuadrado2012,po_2021} and used the generalized gradient approximation (GGA) for the exchange-correlation functional \cite{Perdew1996}.  To evaluate the strain effect, in the case of 0$^\circ$ twist angle,  we considered different configurations $N\times M = (2\times 3)$, $(3\times 4)$, $(4\times 5)$, $(5\times 7)$, and $(7\times 10)$. In these supercells, the calculations are converged for a 350 Ry plane-wave cutoff for the real-space grid with a  $(15 \times 15 \times 1)$  $\vec{k}$-points sampling of the Brillouin zone for the smaller structures, and 15 meV Fermi-Dirac smearing of the electronic occupations. The $\vec{k}$-points is reduced with system size 
down to $(2 \times 2 \times 1)$ mesh for a $10\times 10$ graphene's supercell. We employed a standard exact diagonalization algorithm to compute supercells up to 347 atoms and the pole expansion and selected inversion (PEXSI) \cite{Lin2014} technique, alternative to diagonalization,
to determine the electronic properties and structural relaxation of moir\'e supercells containing up to $\sim$ 1700 atoms. The relaxation was performed in steps, first we relaxed the freestanding strained TMD, then we relaxed the graphene/TMD heterostructure keeping the TMD fixed. We used the conjugate gradient algorithm to minimize the atomic forces below 0.01 eV/{\AA}. The band structures and spin textures were computed from the {\it{ab initio}} Hamiltonian using SISL \cite{zerothi_sisl} as a post-processing tool.
 
\begin{figure}[t]
\centering
\includegraphics[width=\linewidth]{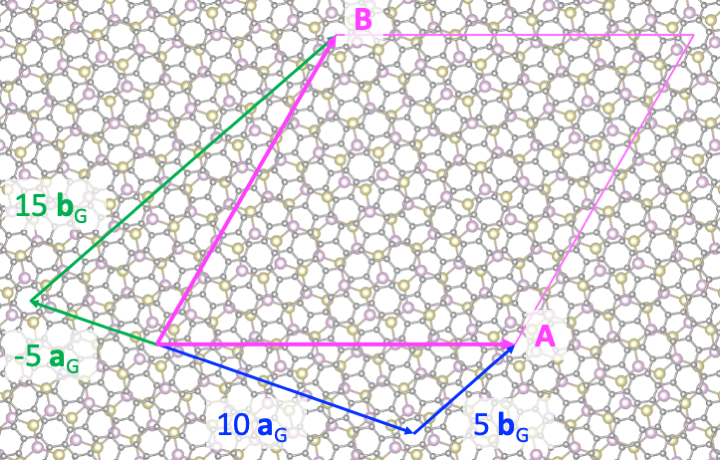}
\caption{The moir\'e periodic cell of
graphene-on-MoTe$_2$, twisted by 30.00$^\circ$. We use linear combinations of the primitive lattice vectors  of graphene $\bm{a}_G$,$\bm{b}_G$ to write the moiré lattice vectors $\bm{A}=10\bm{a}_G+5\bm{b}_G$, $\bm{B}=-5\bm{a}_G+15\bm{b}_G$.}
\label{fig1}
\end{figure}

Figure \ref{fig1} shows an example of the supercell used to define the heterostructure. 
The supercell's lattices vectors ($\bm{A}$, $\bm{B}$) and linear combinations of graphene's lattice vectors ($\bm{a}_G$, $\bm{b}_G$) are shown as magenta, blue and green arrows respectively. Since graphene's lattice vectors form a basis for the supercell, we can use them to write the lattice vectors of the supercells. 
Equivalently, 
the supercell lattice vectors could also be written in terms of TMD's lattice vector. For instance, for a twist angle of $\varphi=30.00^\circ$, we can write $\bm{A}=2\bm{a}_{\rm TMD}+8\bm{b}_{\rm TMD}$ and $\bm{B}=-8\bm{a}_{\rm TMD}+10\bm{b}_{\rm TMD}$ where TMD is MoTe$_2$ in Fig. \ref{fig1}. Both graphene and MoTe$_2$ supercell lattice vectors are shown in Table \ref{twisted_table}. We investigate MoTe$_2$ and WSe$_2$ as representatives of TMDs, and present the corresponding results in the main text and supplementary information \cite{sup_information}. As sulfur and selenium-based TMDs behave similarly \cite{Gmitra2016}, our analysis could be easily generalized to WS$_2$. The relaxed lattice constants of freestanding graphene and MoTe$_2$ are 2.46 {\AA} and 3.55 {\AA} respectively.
 
 \begin{table}[h]
 \hspace{-0.75cm}
 \begin{tabular}{||c|c|c|c||} 
 \hline Twist angle & MoTe$_2$ & Graphene supercell & MoTe$_2$ supercell\\ [0.5ex] 
    ($\theta$ degrees)&  strain{$\%$} &  lattice vectors & lattice vectors  \\ [0.5ex] 
 \hline\hline
 0 & 3.5 &(3,0),(0,3) &(2,0),(0,2) \\ 
 0 ($*a$) & 2.5 &(7,0),(0,7) & (5,0), (0,5)\\ 
 0 & 2.0 &(7,0),(0,7) & (5,0), (0,5)\\ 
 0 & 0.5 &(10,0),(0,10) & (7,0), (0,7)\\ 
 0 ($*b$) & -0.71 &(10,0),(0,10) & (7,0), (0,7)\\ 
15.21 & 0.61 &(8,7),(-7,15)&(3,7),(-7,10) \\ 
15.00 & 1.95 &(16,9),(-9,25) & (7,10), (-10,17) \\ 
15.75 & 2.45 &(18,7),(-7,25) & (15,0),(0,15) \\ 
14.99 & 3.55 &(14,3),(-3,17) & (7,5),(-5,12) \\ 
30.00 & -0.71 &(10,5),(-5,15) & (2,8), (-8,10) \\
30.49 & 2.35 &(13,5),(-5,18) & (3,9),(-9,12)\\
30.00 & 3.26 &(13,0),(0,13) &(5,5),(-5,10) \\
 \hline
 \end{tabular}
       \caption{\ Twist angle $\theta$, strain, and supercell lattice vectors of graphene/MoTe$_2$ heterobilayers. The supercell lattice vectors are given in terms of the primitive graphene and MoTe$_2$ unit cells lattice vectors rotated by $\varphi$ degrees with respect to each other. The relaxed lattice constant for graphene is $a_{\rm gr}= 2.46$ {\AA} while the TMD lattice constant $a_{\rm MoTe_2}$ changes accordingly the values presented in the table above.(*) In these cases, graphene's lattice constant was slightly modified to 2.42 {\AA} (a) and 2.47 {\AA} (b).}
  \label{twisted_table}
\end{table}

\begin{figure}[h]
\hspace{-1.cm}
\centering
\includegraphics[width=1.1\linewidth]{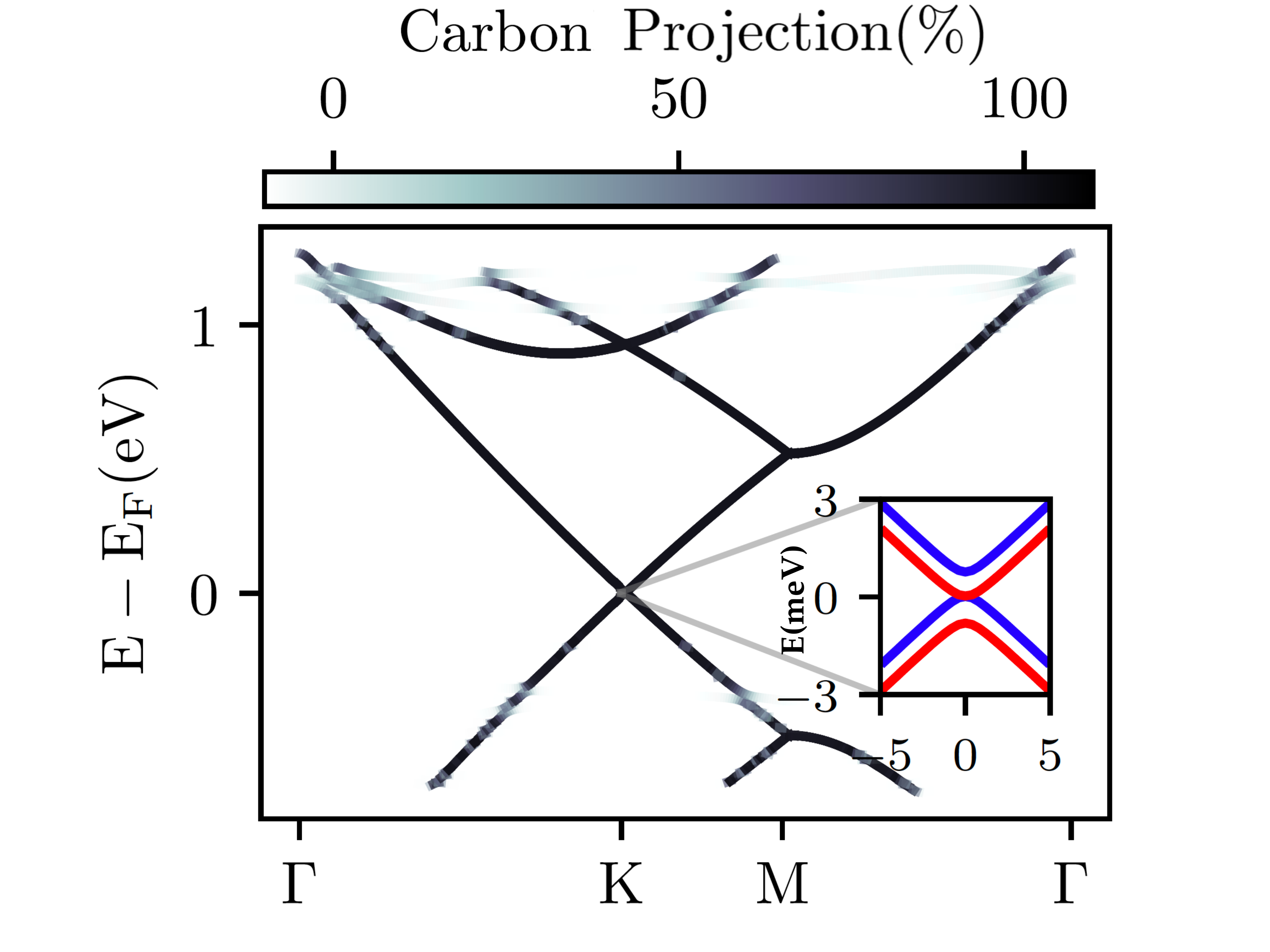}
\caption{ Projected band structure of graphene on monolayer MoTe$_2$ at zero twist angle. The gray scale indicates the projection of the band-states into carbon orbitals. The inset shows the band structure around the Fermi level where the colors represents bands with opposite spins. These results were obtained in a $7\times7$ graphene supercell on a $5\times 5$ MoTe$_2$ supercell. }
\label{fig2}
\end{figure}

 The band structure of graphene/MoTe$_2$ heterostructure with $0.5\%$ strain is presented in Fig.~\ref{fig2}. The grayscale indicates the projection of the band states into the carbon orbitals and demonstrates that graphene's Dirac cone is preserved and lies within the MoTe$_2$ bandgap, maintaining its $p_z$ nature. This result justifies the use of an effective graphene-only Hamiltonian in subsequent sections. The inset represents a zoom within a -3 to 3 meV energy window, where there is an evident spin splitting of $\sim0.5$ meV, which we attribute to the proximity-induced SOC. Our results differ from a previous work where the position of the Dirac cone is below the valence band of the MoTe$_2$ \cite{Gmitra2016}, because in this latter work supercells with strain values higher than $10\%$ were used producing an artificial enhancement of charge transfer that shifted the Dirac's cone position \cite{romero_mos2_vdw, Di_Felice_2017}.

 Free standing graphene belongs to the $D_{6h}$ point group that only allows for the presence of a Kane-Mele type SOC \cite{KaneMelePRLQSH}. On top of a TMD substrate, graphene's symmetry is reduced to $C_{3v}$ dictating the following Hamiltonian \cite{KochanPRB2017}:
\begin{equation}
H=\hbar v_{\rm F}(\tau \sigma_xk_x+\sigma_yk_y)+\Delta \sigma_z +H_{\rm soc},
\end{equation}
where $\hbar$ the reduced Planck's constant, $v_{\rm F}$ the Fermi velocity, $\mathbf{k}=(k_x,k_y)$ the crystal momentum around the Dirac points, $\tau$ is $1/-1$ for the K/K' Dirac's cones, $\sigma_i$ (with $i=x,y,z$) the Pauli matrices acting on the sublattice space. The second term is a staggered potential that opens a bandgap of size $\Delta$. Finally, the third contribution  
\begin{equation}
H_{\rm soc}=\lambda_{\rm I}\sigma_z \tau_z s_z+ \lambda_{\rm VZ}\tau_z s_z+\lambda_{\rm R}(\kappa \sigma_x s_y-\sigma_y s_x)
\end{equation}
models the spin-orbit coupling, and consists of the intrinsic, valley Zeeman (VZ), and Rashba spin-orbit couplings with strength $\lambda_{\rm I}$, $\lambda_{\rm VZ}$, and $\lambda_{\rm R}$ respectively, where $s_i$ (with $i=x,y,z$) represents the Pauli matrices acting on the spin space \cite{Gmitra2016,KochanPRB2017}. It is important to higlight that despite the presence of uniform strain and twist angle, the point group is unchanged given that both structures belong to $C_{3v}$ in this configuration. However, for cases with uniaxial strain or different structural phases of the TMD new spin-orbit manifestations could be necessary.

\section{Twist-angle and strain}

\begin{figure}[!htb]
\hspace{-1.cm}
\centering
\includegraphics[width=1.1\linewidth]{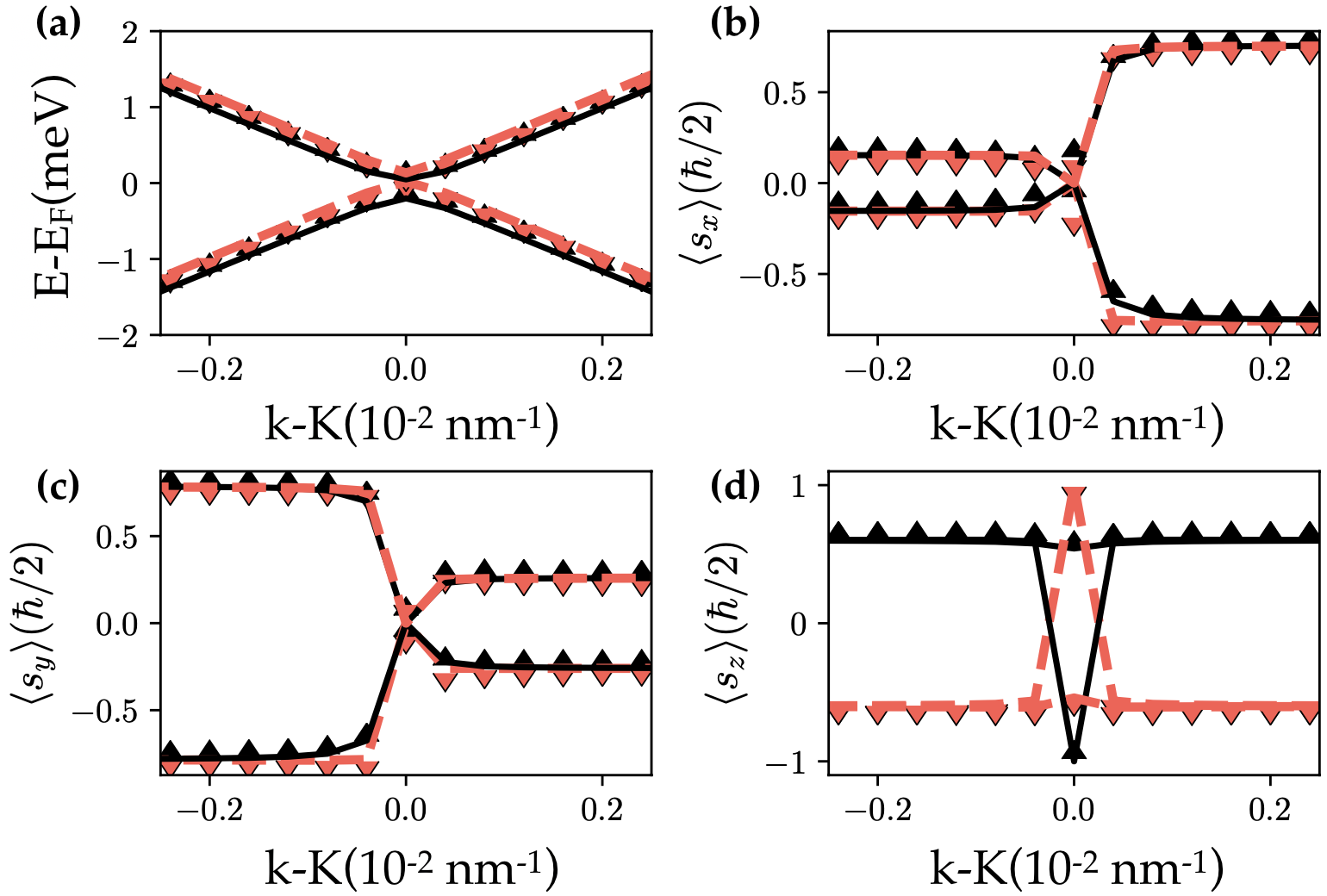}
\caption{Electronic band structure around one of the Dirac points (a) and spin texture components (b-d) computed from DFT (symbols) and model Hamiltonian (lines) for graphene/MoTe$_2$ twisted by 15$^{\circ}$. Red and black colors identify spin up and down, respectively. Spin textures in momentum space are computed along the same path as the band structure.}
\label{fig3}
\end{figure}

In Fig.~\ref{fig3}, we show the band structure and spin texture around the Dirac point for the heterobilayer twisted by a $15^{\circ}$ angle. We found a remarkable agreement between our model (lines) and DFT calculations (symbols). This result shows that the $C_{3v}$ symmetry-based model is sufficient to capture the relevant physics of the twisted system and yields to similar results obtained via Slater-Koster methods \cite{Alsharari2018,Li2019,David2019}. Moreover, we see a drastic change in the SOC splitting with respect to the untwisted case, displaying a reduction from $\sim  0.5$ meV to values below 0.1 meV. These results are the first demonstration via DFT that twisting can strongly modulate proximity interaction and, hence, SOC effects. Following a similar procedure, we determined the spin-orbit coupling parameters at different strain and twist angle values shown in Table \ref{twisted_table}. To separate the role of strain and twist angle we choose supercells with similar strain values evaluating in each case the layers misalignment effects.

\begin{figure}[htp]
  \includegraphics[clip,width=1.1\columnwidth]{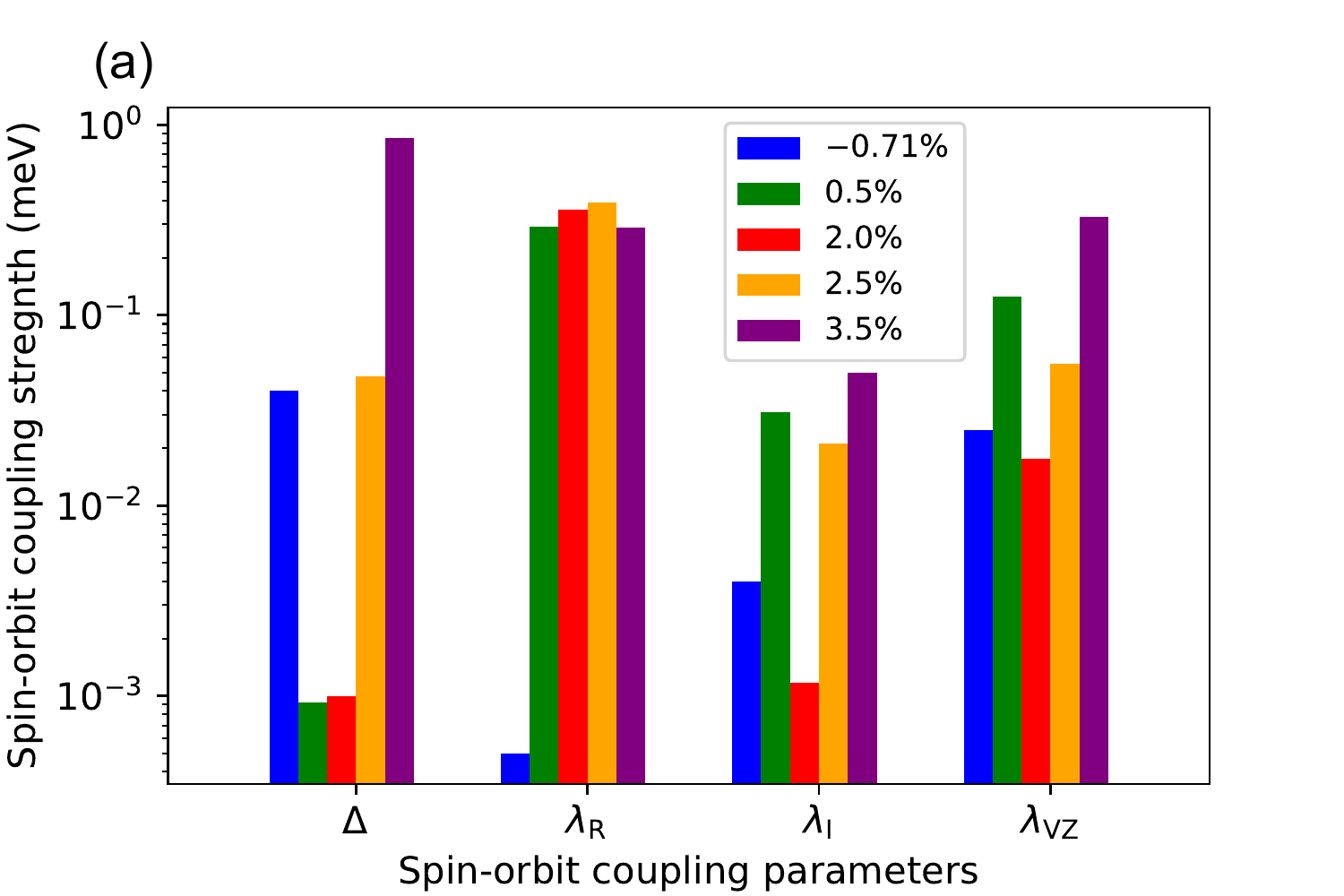}

  \includegraphics[clip,width=1.1\columnwidth]{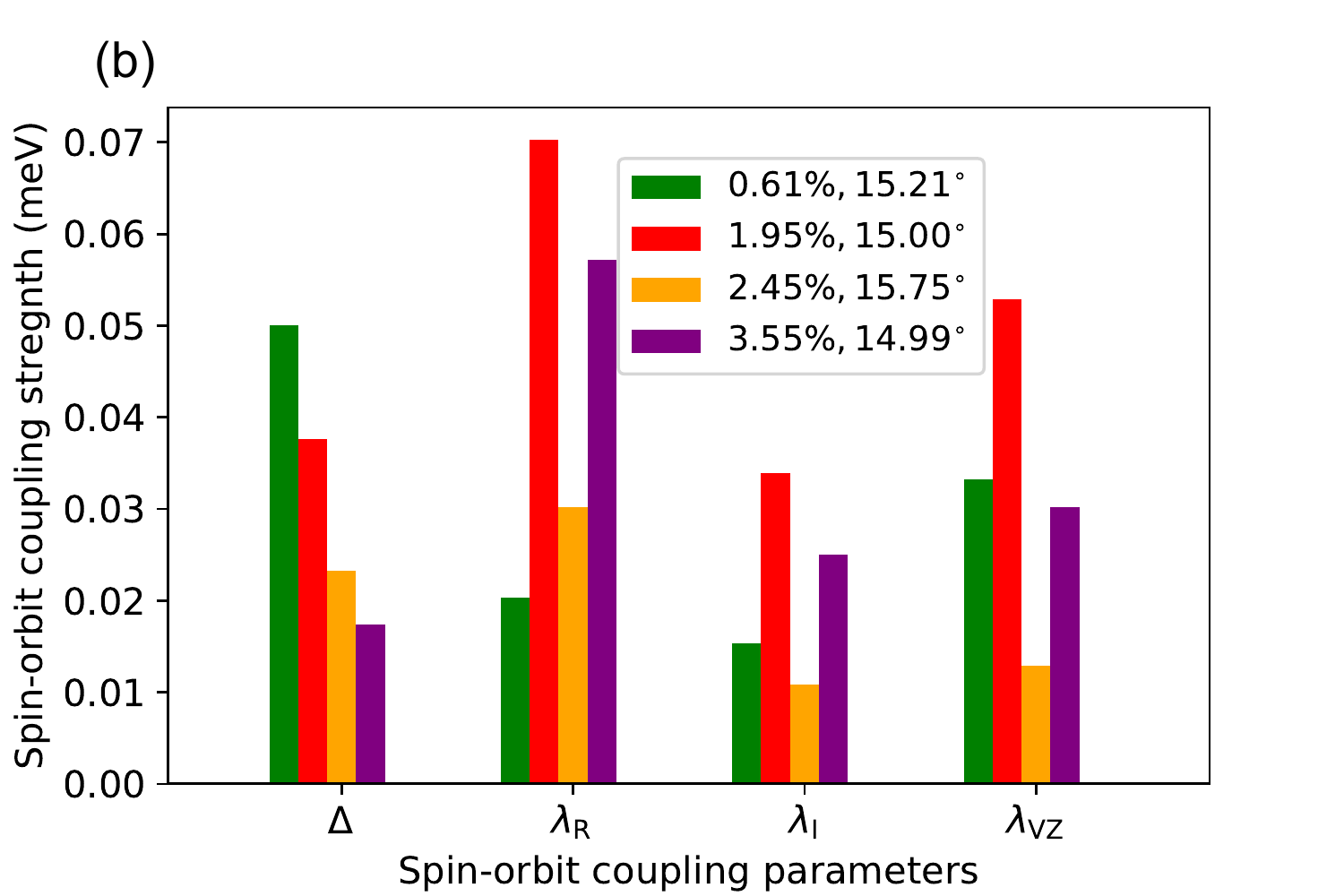}%

  \includegraphics[clip,width=1.1\columnwidth]{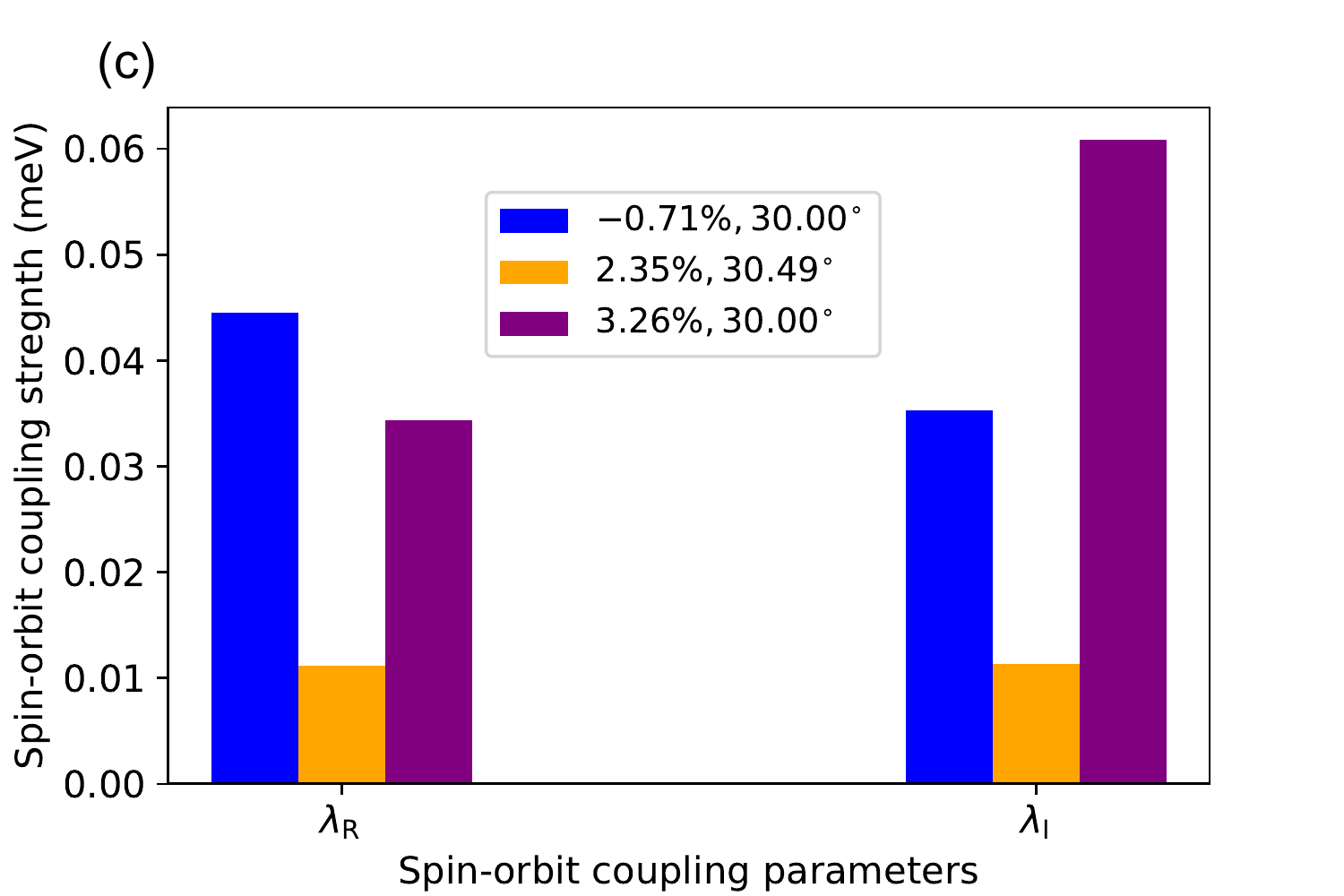}
\caption{Band gap ($\Delta$) and SOC parameters extracted for graphene/MoTe$_2$ at different percentage of strain and angles. In (a) we show the untwisted case for 0$^\circ$, (b) displays the values for angles $\sim$ 15$^\circ$ while in (c) there are the angles $\sim$ 30$^\circ$. Color code is indicated in the inset. A logarithmic scale is used only in the case of zero twist angle.}\label{fig4}

\end{figure}

In Fig.~\ref{fig4}.a, we present the results for zero twist angle heterostructures with different strains. The main outcome is a dominant and mostly strain-independent Rashba SOC having the largest values of $\lambda_R$  with strain up to $3.5\%$, except for the negative strain $-0.71\%$, where the VZ interaction exceeds the other contributions. We suggest that this behavior is related to the Dirac cone position within the TMD band gap. For instance, the structures with $-0.71\%$ and $3.5\%$ strain present larger $\lambda_{VZ}$ values when compared to the Rashba parameters (note the log scale); this is corroborated by their band structures depicted in the supplementary material (\cite{sup_information}), where the Dirac cone position lies closer to the valence band, hence, resembling to the situation for MoS$_2$ where the out-of-plane component of the spin texture is the largest one. We reached this conclusion after considering the expression for the VZ interaction, described by the following equation \cite{David2019}
\begin{equation}
    \lambda_{\rm VZ} \hspace{0.25cm}\propto \hspace{0.25cm}\sum_{b_i}\frac{\Delta_{s}}{\Delta E^2-\Delta^2_{s}},
    \label{eq_vz}
\end{equation}
\noindent
where $\Delta E$ is the energy difference between TMD bands (b$_i$) and the Dirac point without taking into account SOC, $\Delta_{s}$ is the spin splitting of the TMD bands due to the intrinsic SOC. In this case, from Eq. \ref{eq_vz}, we infer that the position of the Dirac cone is a signature of a system with a small VZ interaction which might be the result of the twisting and electrostatic nature of the interfacial interaction. The Rashba interaction takes a more complicated form but remains mostly insensitive to the relative position of the Dirac cone and depends more on the strength of the spin-orbit coupling of the TMD's constituent atoms, explaining a larger Rashba obtained for MoTe$_2$ compared to WSe$_2$ as a result from the difference between Te and Se atomic numbers. Additionally, we see a strong modulation of the VZ interaction with strain, in particular, VZ is generally smaller than Rashba for low strain, a situation that contrasts with previous calculations performed considering different TMDs \cite{Gmitra2016,AlsharariPRB2016,Alsharari2018,Li2019,David2019}. Recent studies \cite{Li2019,David2019} suggest tunneling as the origin of both Rashba and Valley-Zeeman SOC. Therefore, a given condition that minimizes the Rashba SOC should suppress VZ simultaneously. Moreover, the VZ interaction is predicted to peak for certain angles when the Dirac cone lies close to the conduction band \cite{David2019}. In the MoTe$_2$-based heterostructures, the Dirac cone lies instead near the valence band, so we expect VZ to be maximal for the untwisted structure \cite{David2019},  which is in agreement with our results. We note that there is currently no theory that supports our observed behavior, where Rashba SOC dominates over all other interactions. Such a regime is said to display optimal spin-to-charge conversion \cite{Offidani2017} and unconventional spin-orbit torque \cite{Sousa2020} which is expected to occur in pristine graphene for instance, although there is still no transport evidence demonstrating it \cite{Benitez2018}. We attribute this hitherto unknown regime to the difference in lattice constant of graphene and MoTe$_2$. Selenium- and sulfur-based TMDs possess similar lattice constants and display similar moir\'e patterns. The lattice constant of tellurium-based TMDs is 10\% larger and changes the periodicity of the moir\'e pattern. An indication of this behavior is the fact that for graphene on MoTe$_2$ we need to build a much larger supercell (10$\times$10 graphene supercell) compared to the other TMDs (4$\times$4 graphene supercell) to attain a similar level of strain when studying the untwisted structures. 
The 3.5$\%$ strain case displays a different behavior than its less strained counterparts. We traced back this phenomenon to a K\'ekule distortion appearing in this particular arrangement due to the fact that there is a metallic and a chalcogen atom at the center of two different carbon rings, making them nonequivalent and distorting the nearby hoppings \cite{sup_information}. Another evidence of the K\'ekule distortion is the large band gap observed for this structure. This situation resembles early calculations of graphene on top of topological insulators \cite{Song2018}. In this work, we omitted the SOC parameters corresponding to the 3 (TMD)$\times$4 (graphene) supercell due to its value of strain larger than 5$\%$, which buries the Dirac cone into the TMD valence bands, preventing the use of a graphene-only hamiltonian. 

We will now discuss the effect of twist angle and its interplay with strain. In Fig. \ref{fig4}.b we show the SOC parameters obtained for heterobilayers rotated by an angle $\theta=15^\circ$. The most prominent result is a strong reduction of all parameters compared with the untwisted structures, which indicates a suppression of proximity-effects. We here argue that this behaviour can arise from the modified distances between Tellurium and Carbon atoms due the relative rotation between the layers\footnote{The interlayer distance has remained almost the same, here we refer to the change in the in-plane positions after rotating one layer with respect to the other.}. 
However, the general trend is in agreement with previous calculations that attribute it to a change in the tunneling parameters due to an indirect coupling between the graphene and TMD bands \cite{David2019}. We clearly identify the predominance of Rashba SOC except for the smallest strain 0.61$\%$, where $\lambda_{VZ}$ is larger. The larger values of $\lambda_R$ are expected from symmetry considerations since the horizontal mirror plane is lifted at any angle. The staggered potential and VZ interaction are both associated with a vertical mirror plane that can be restored at certain twist angles. 

\begin{figure*}[ht]
\centering
\includegraphics[scale=0.35]{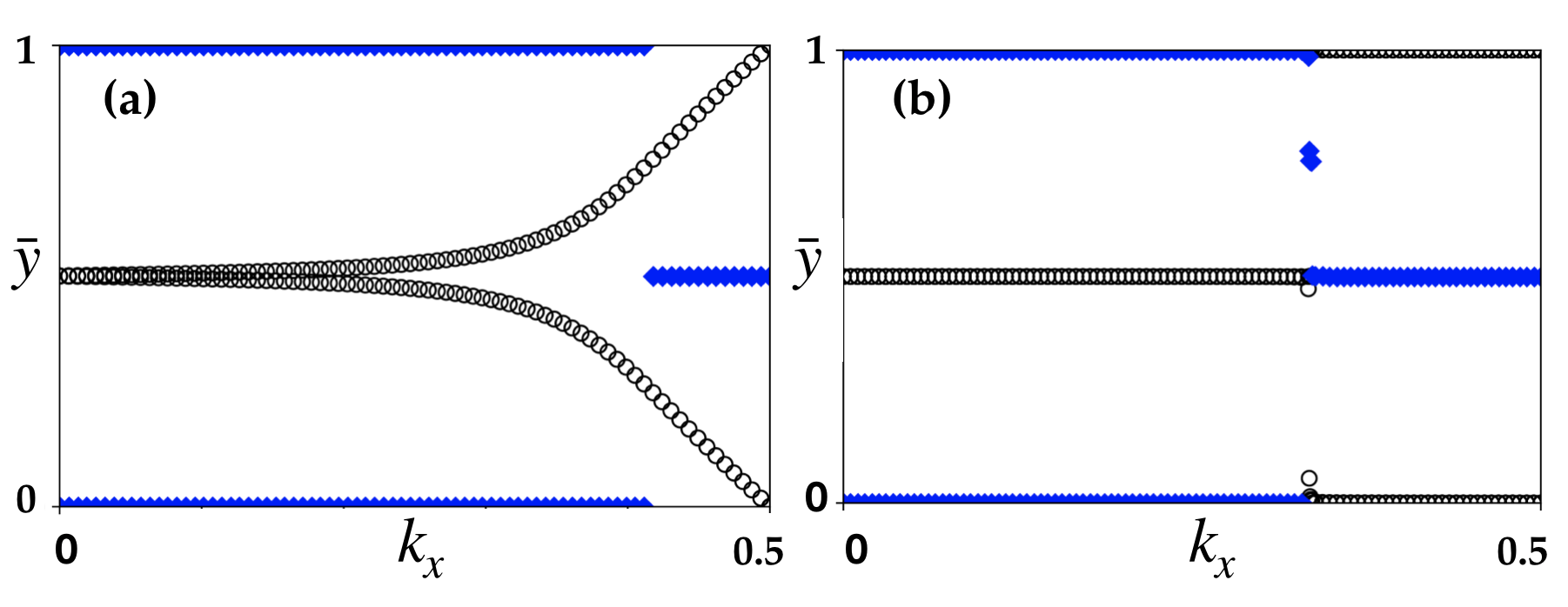}
\caption{$\mathbf{Z}_2$ Topological invariant by means of the Wannier charge centers (WCC's) for (a) WSe$_2$/graphene ($\nu=1$) and (b) MoTe$_2$/graphene ($\nu=0$) heterostructures. Circles represent the evolution of the WCC's within a unit cell as a function of $k_x$ while blue rhomboids mark midpoints between largest gaps. The appearance of two disjoint diamonds in (b) suggest an even number of crossings with the WCC's evolution, characterizing a trivial topological phase.}
\label{topo_invariant_hetero}
\end{figure*}

 In Fig.~\ref{fig4}.c we extend our analysis to the bilayer with a twist angle of  $\theta=30^\circ$. We omit in the plot the band gap $\Delta$ and $\lambda_{VZ}$ parameters as they become negligible on the scale used to show $\lambda_R$ and $\lambda_I$. The overall conclusion is that rotating the structures with respect to each other, suppresses the tunneling between monolayers reducing the SOC parameters, such that they attain their smallest values for the 30$^\circ$ twist angle. This could explain the variability in the SOC parameters measured over different experiments, growing conditions, and substrate choice \cite{PhysRevLett.121.127703,Ghiasi2017}. Under the same assumption based on Eq. \ref{eq_vz}, the physical origin of the reduction in the parameters can be related to the graphene's Brillouin Zone rotation, leading to a decreasing in the tunneling between carbon p$_z$ orbitals and the TMD bands as we depict in \cite{sup_information}. From the two non-vanishing parameters in this regime we note modulations in $\lambda_R$ and $\lambda_I$ for different strain values. This leads us to point out the sensitivity to strain effects in these type of heterostructures, which might have an impact in experimental measurements.
 
 We also found a topological transition in the heterostructure by calculating the $\mathbf{Z}_2$ topological invariant depicted in Fig. \ref{topo_invariant_hetero}. We calculated the topological index by means of $\mathbf{Z}2$Pack \cite{z2pack}. From the tight-binding hamiltonian we can obtain the hybrid Wannier charge centers (WCC's) $\bar{y}$'s defined modulo a lattice vector due to the ambiguity in determining the unit cell \cite{SPALDIN20122,hwcc}, in this sense, rather than computing the Chern number as an integral of the Berry curvature over the Brillouin zone, the Chern number $C$ is calculated like
\begin{equation*}
    C=\frac{1}{a_y}\left(\sum_n \bar{y}_n(k_x=2\pi)-\sum_n \bar{y}_n(k_x=0)\right),
\end{equation*}
\noindent
where $\bar{y}$'s are assumed to be smooth functions in $k_x$. The fact that these functions are periodic modulo a lattice vector $a_y$, means that a non-zero Chern number is related to the impossibility of finding a smooth and periodic set of Bloch functions for the those bands. When imposing Time reversal symmetry (TRS), the topological classification is given in terms of the $\nu$ ($\mathbf{Z}_2$) invariant. In this case $S_1$ and $S_2$ \footnote{Recall that we are considering the more general case when spin mixing terms are allowed such that the Spin is not a valid good quantum number anymore.} subspaces have states related by TRS which carry opposite Chern numbers, therefore the topological invariant $\nu$ will be either zero or one
\begin{equation*}
    \nu =(C_{S_1}-C_{S_2})/2 \hspace{.15 cm} \mod \hspace{0.15 cm} 2,
\end{equation*}
\noindent
given that for non-trivial topological phases $C_{S_{1,2}}$ are odd. For the MoTe$_2$-based heterostructure Fig. \ref{topo_invariant_hetero}(b) we observed a trivial phase for all values of twist angle, but for the 
WSe$_2$ heterostructure Fig. \ref{topo_invariant_hetero}(a) it exhibits a topological transition from a nontrivial state in 15$^\circ$ to a trivial in 30$^\circ$.  For the latter heterostructure the twist angle is able to modulate the Kane-Mele interaction by an order of magnitude, activating the topological transition. This represents a plausible explanation for the unusual large topological behavior measured recently via weak antilocalization measurements \cite{WakamuraTPRL2018,WakamuraTPRB2019}, opening a door for tuning topological properties via twist angle.

\section{Spin dynamics}

Since our model can well reproduce the spin-textures for all our configurations, it can be used to determine the spin-lifetime anisotropy within the Dyakonov-Perel regime \cite{Fabian2007,AronPRL2017,Song2018,LSQT2020}. To show this, let us recall that the electron's spin under the effect of an applied magnetic field will undergoes Larmor's precession with a frequency $\omega= \Delta_Z/\hbar$, where $\Delta_{Z}$ the Zeeman splitting. In the Dyakonov-Perel regime, the SOC can be considered as a momentum-dependent magnetic field given by $\bm{B}(\bm{k})=\omega \langle \bm{s}\rangle_{\bm{k}}$, with  $\langle \bm{s}\rangle_{\bm{k}}$ the spin texture \cite{Song2018}. Since the direction of this field depends on the momentum,  scattering processes will induce random fluctuations  that will lead to spin relaxation at the following rate

\begin{equation}
\tau_{\rm s,\alpha}^{-1} = \omega \tau_{\rm *, \alpha}( |\bm{S}|^2 -S_{\alpha}^2), \end{equation}

\begin{figure}[ht]
\hspace{-1.25cm}
\includegraphics[width=1.1\linewidth]{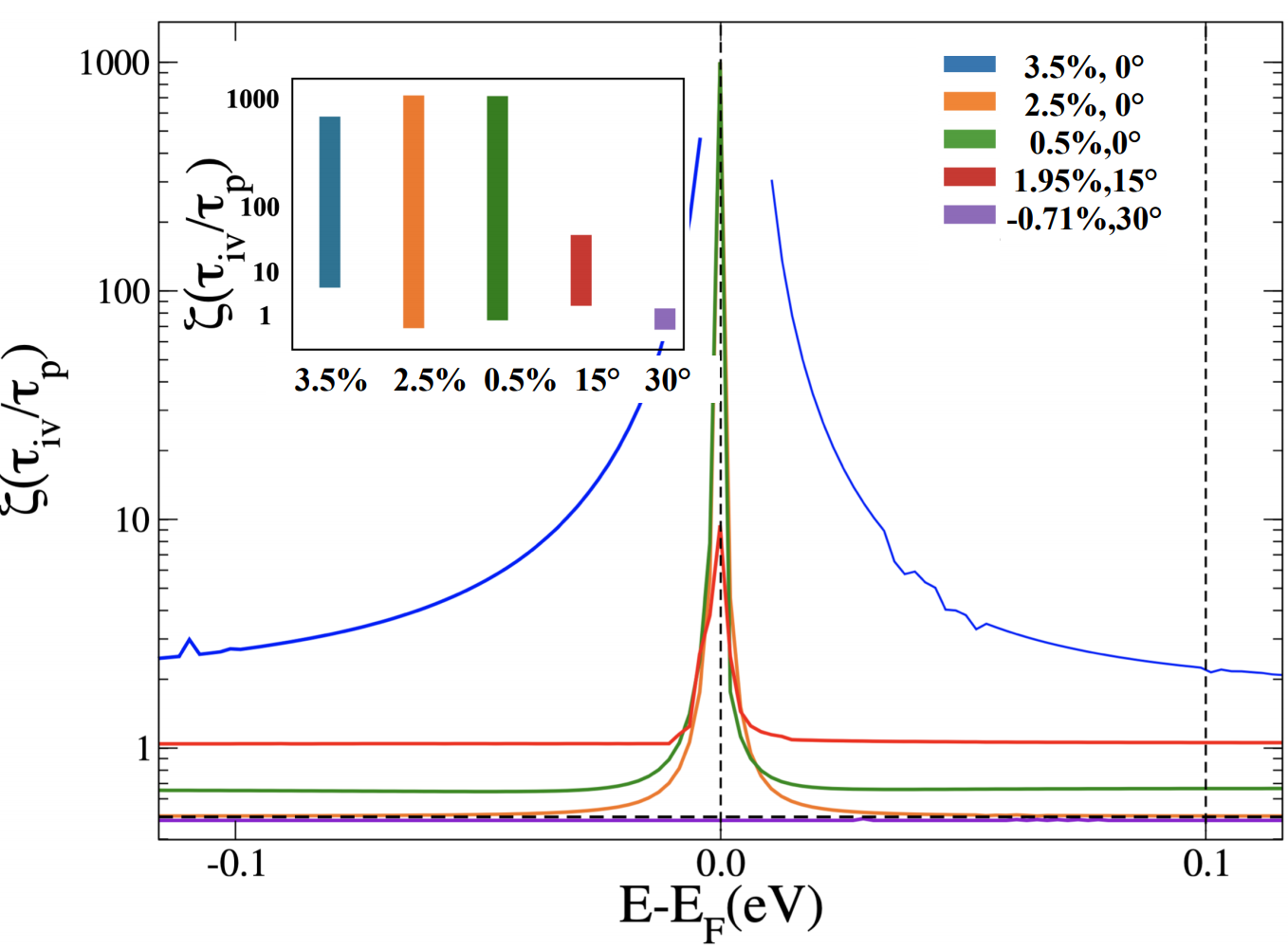}
\caption{Spin-lifetime anisotropy in terms of $\tau_{iv}/\tau_{p}$ for different structures. We can see how the anisotropy increases with valley Zeeman. The vertical dashed lines represent the energy values at which we identified the maximum and minimum anisotropy values. The inset shows the range of values taken by the anisotropy for each case. We label the moiré structures by their twist angle. The smallest anisotropy range is obtained for the 30$^\circ$ twisted heterostructure.}
\label{fig5_anisotropy}
\end{figure}

\noindent
where $S_\alpha^2 \equiv \langle s_\alpha\rangle^2$ the average of the $\alpha=x,y,z$ component of the spin texture around the Fermi level, and $\tau_{\rm *, \alpha}$ the time that takes to randomize the $\alpha$-component of the effective magnetic field \cite{Fabian2007}. It is usual to consider $\tau_{\rm *, x}=\tau_{\rm *, y}=\tau_{\rm *, z}=\tau_{\rm p}$, with $\tau_{\rm p}$ the momentum relaxation time. But graphene's special SOC allows for an extra contribution to in-plane rates when there is an out-of-plane component of the spin texture. Due to time-reversal symmetry, such a texture leads to spin-valley coupling, which activates the relaxation via intervalley scattering processes characterized by the time $\tau_{\rm iv}$ \cite{AronPRL2017}. Accordingly, by using 
Mathiessen's Rule one gets
\begin{equation}
    \tau_{\rm *,x}^{-1}=\tau_{\rm *, y}^{-1}=\tau_{\rm iv}^{-1}+\tau_{\rm p}^{-1}.
    \label{rates}
\end{equation}

The spin-lifetime anisotropy $\zeta$, is defined as the ratio between out-of-plane and in-plane spin relaxation times $\zeta\equiv \tau_{z}/\tau_{x}$ \cite{Raes2016}, and following the discussion above, it is easy to see that it is directly related to the spin textures \cite{Song2018}
\begin{equation}
\zeta=\frac{\tau_{z}}{\tau_{x}} = \frac{1}{1+\tau_{\rm p}/\tau_{\rm iv}} {\sum_{n} \bm{S}_n^2-S_{x,n}^2\over \sum_n \bm{S}_n^2-S_{z,n}^2}\label{anisotropyEq}, 
\end{equation}
where the sum is over all bands crossing the Fermi level.
The spin-lifetime anisotropies computed by applying Eq.~\ref{anisotropyEq} to the different configurations 
are presented in Fig.~\ref{fig5_anisotropy}. We note that the largest modulations in SOC happens for the zero twist angle, we have chosen the set of values below as representative of all the parameters considered in this work, namely strain and twist angle. The most prominent result is a giant spin-lifetime anisotropy at the Dirac point that is a consequence of the combination of the Rashba SOC and the staggered potential. Let us remind that due to spin-pseudospin coupling \cite{Tuan2014}, the staggered potential will polarize both pseudospin and spin graphene's degrees of freedom out of the plane close to the Dirac point, creating a texture similar to that one produced by valley-Zeeman interaction \cite{GmitraPRL2013,garcia2018csr,Zollner2020}. At higher energies, the anisotropy decays because the bands become linear, and the texture becomes solely in-plane. As discussed previously, in-plane spin textures allow to consider all relaxation rates equal to the momentum relaxation time and lead to an anistropy of $\zeta=1/2$, which is consistent with our results.  The effect of the gap-induced spin-valley coupling is more prominent in the case of the distorted K\'ekule structure because of its giant gap and is similar to previous results of graphene on Bi$_2$Se$_3$ \cite{Song2018}. These results highlight two things, first that a giant spin-lifetime anisotropy is not a smoking gun of valley-Zeeman interaction but of spin-valley coupling, and a gate-dependence should always be performed to identify the different mechanisms, and also that rotation angle may induce substantial changes in the spin and transport properties of graphene/TMD heterostructures.

\section{Conclusions}

We have found that the spin-orbit coupling of graphene/TMD heterostructures can be strongly modulated by strain and layers misalignment. Our results for the WSe$_2$ are in agreement with previous tight-binding-based calculations \cite{Alsharari2018,Li2019,David2019}, but we found a strong dependence of the intrinsic SOC coupling that provokes a topological transition. Additionally, the prediction of giant enhancement for SOC parameters, reaching values as large as 3.0 meV \cite{Alsharari2018,Li2019,David2019}, is not supported by our DFT results. This is particularly true for MoTe$_2$, where graphene Dirac cone is very close to the valence band of MoTe$_2$, and accordingly, it is expected to present a massive imprinted spin-orbit coupling. By an extensive study considering different misalignements and strain values, we found that the Rashba SOC is the most sensitive to strain and twisting angle, enabling us to study the interplay with other SOC interactions and allowing us to tune the system into an ideal Dirac-Rashba model, as a result of particular interest for low-power memories based on spin-orbit torque \cite{RMPManchon_2019}. We also show that a gate-dependent giant spin-lifetime anisotropy would be conclusive evidence of such state. These findings highlight the impact of orientation angle between layers, which could serve as guidance for angle-engineering that will aim at maximizing the proximity-induced SOC and spin transport figure of merits. 

\section{acknowledgments}

AP was supported in part by the Coordenação de Aperfeiçoamento de Pessoal de Nível Superior - Brasil (CAPES) - Finance Code 001. AF was supported by  FAPESP (Grant 16/14011-2), JHG and SR 
were supported by the European Union Horizon 2020 research and innovation programme under Grant Agreement No. 881603 (Graphene Flagship).
ICN2 is funded by the CERCA Programme/Generalitat de Catalunya (Grant 2017SGR1506),
and is supported by the Severo Ochoa program from Spanish MINECO (Grant No. SEV-2017-0706).
ZZ acknowledges financial support by the Ramon y Cajal program RYC‐2016‐19344 (MINECO/AEI/FSE, UE). ZZ, NW, and PO acknowledge support from the EC H2020-INFRAEDI-2018-2020 MaX "Materials Design at the Exascale" CoE (grant No. 824143) and Spanish MCIU/AEI(grant No. PGC2018-096055-B-C43). ZZ acknowledges the Netherlands Sector Plan program 2019-2023. NW has received funding from the European Union's Horizon 2020 research and innovation programme under the Marie Skłodowska-Curie grant agreement No. 754558.
ZZ and NW acknowledge
the computer resources at MareNostrum and the technical support provided by Barcelona Supercom-
puting Center (FI-2020-1-0014). We acknowledge PRACE for awarding us access to MareNostrum 4 at Barcelona Supercomputing Center (BSC), Spain (OptoSpin project id. 2020225411).


\bibliography{mybib}

\end{document}